1    **Guillaume KON KAM KING**

2    **Address:** UMR CNRS 5558 – LBBE "Biométrie et Biologie Évolutive" UCBLyon 1

3    43 bd du 11 Novembre 1918

4    69622 VILLEURBANNE cedex, France

5    **Phone:** +33 4 26 23 44 71

6    **Email:** guillaume.kon-kam-king@univ-lyon1.fr

8    **Number of words:** 5647

9    **Number of references:** 40

10    **Number of tables:** 0

11    **Number of figure legends:** 2

# MOSAIC_SSD: a new web-tool for the Species Sensitivity Distribution, allowing to include censored data by maximum likelihood

**Guillaume Kon Kam King†, Philippe Veber,† Sandrine Charles,†,‡ and Marie Laure Delignette-Muller†,§**

†UMR CNRS 5558 – LBBE "Biométrie et Biologie Évolutive" UCBLyon 1

43 bd du 11 Novembre 1918

69622 VILLEURBANNE cedex, France

‡ Institut Universitaire de France, 103 bd Saint-Michel, 75005 Paris, France

§ VetAgro Sup Campus Vétérinaire de Lyon 1, avenue Bourgelat 69280 MARCY L'ÉTOILE

35

36

37

38

39

40

41

42

43

44

45

46

47

48

49

50

51

52

53

54

55

56

Corresponding author : sandrine.charles@univ-lyon1.fr

57

## Abstract

Censored data are seldom taken into account in Species Sensitivity Distribution (SSD) analysis. However, they are found in virtually every dataset and sometimes represent the better part of the data. Stringent recommendations on data quality often entail discarding a lot of this meaningful data, often resulting in datasets of reduced size, which lack representativeness of any realistic community. However, it is reasonably simple to include censored data into SSD by using an extension of the standard maximum likelihood method. In this paper, we detailed this approach based on the use of the R-package *fitdistrplus*, dedicated to the fit of parametric probability distributions. We especially present the new web-tool MOSAIC_SSD that can fit an SSD on datasets containing any type of data, censored or not. MOSAIC_SSD predicts any Hazardous Concentration (HC) and provides in addition bootstrap confidence intervals on the predictions. Finally, we illustrate the added value of including censored data in SSD taking examples from published data.

69

**Keywords**: bootstrap, confidence interval, web interface, fitdistrplus, hazardous concentration

71

72

73

74

75

76

77

78

79

80

## Introduction

The Species Sensitivity Distribution (SSD) is a central tool for risk assessment in ecotoxicology. It provides a subtle way to define hazardous concentrations for the environment compared to using arbitrary safety factors. However, the approach is still subject to heated debates regarding its optimal implementation where no consensus has been reached so far. As a result, environmental regulatory bodies often advocate country or region-specific approaches[1–5]. In studying SSD, methodological choices based on very theoretical arguments have a direct impact on legislation and significant economic and ecological outcomes. The purposes of the SSD approach are to model interspecies sensitivity variability and to provide a protective concentration for a group of species. SSD uses sensitivity data as input, such as No Effect Concentration (NEC), No Observed Effect Concentration (NOEC) or Effective Concentration for $x$% ($\mathrm{EC}_x$) for a group of tested species. These can be coined Critical Effect Concentrations (CECs), which are obtained through acute or chronic toxicity bioassays. The SSD approach is based on the hypothesis that variability in species sensitivity follows a probability distribution. This distribution is extrapolated from the sample of tested species to infer a group-wide protective concentration, the *Hazardous Concentration* for $p$% of the group ($\mathrm{HC}_p$). In general, a parametric distribution is assumed. However, assumptions about the distribution can be avoided using non-parametric methods[6].

The SSD approach has many issues[7] ranging from ecological to statistical concerns. Despite the difficulty of addressing all of them, specific aspects can be improved. Throughout the body of work mentioning SSD, there are very few occurrences of taking into account properly missing data, non-detects or censored data in general. On those rare occurrences, this is

103 achieved in a Bayesian framework[8], which requires a statistical expertise not accessible to

104 untrained users. However, censored data contain crucial information and ignoring or

105 transforming them alters the quality of the predictions based on the remaining data[9]. It is

106 possible to deal with censored data in a more familiar frequentist framework using standard

107 maximum likelihood methods. This also offers several advantages over common SSD

108 approaches. In this article, we present a web-tool, MOSAIC_SSD (http://pbil.univ-

109 lyon1.fr/software/mosaic/ssd/), to easily include censored data in a SSD analysis. MOSAIC_SSD

110 relies on the existing R package, *fitdistrplus[10]*.

111  In this paper, we first reviewed the methods to fit a distribution to data, in order to

112 position our tool among the existing SSD approaches. Then, we explained how to take censored

113 data into account and presented the web-tool we have designed for this purpose, MOSAIC_SSD.

114 Finally, we illustrated the added value of including censored data in SSD using example datasets

115 from published literature.

116

## 117 Review of methods to fit an SSD

118

119  Two steps are required to fit an SSD. The first step is to choose a distribution which

120 seems appropriate to describe the data. Possible options include 1) a Weibull distribution[11] to

121 emphasize the tails of the distribution, 2) a triangular distribution[11] with a finite-size support

122 when no species are more sensitive/resilient than a certain threshold value, 3) a multimodal

123 distribution[12,13] for an assemblage of several taxons, etc. Log-normal[14] and log-logistic[15]

124 distributions are the customary choices[16] although an extensive list of distributions have been

125   applied to SSD. Alternatively, distribution-free methods can be used to avoid the subjective

126   choice of a distribution for the data. This approach has been used for SSD in various works

127   [6,8,17]. Among these distribution-free methods, the Kaplan-Meier estimator is a means to

128   include censored data in SSD[8]. However, it is restricted to certain types of censored data. One

129   difference between parametric and distribution-free methods is the following: when fitting a

130   parametric distribution, the possibilities of shapes are restricted to a certain class of distributions

131   and the shape is refined by adding information from the data. In a distribution-free approach, any

132   kind of shape is allowed and the result emerge solely from the data. Therefore, distribution-free

133   methods do not use any sort of exterior information and often require more data than parametric

134   methods[16,17]. Consequently, when dealing with small datasets it is more reasonable to fit a

135   parametric distribution.

136   Each parametric distribution is determined by a set of parameters, so the second step is to

137   estimate them. Parameter estimation is performed by optimizing a chosen criterion: the

138   likelihood or some goodness-of-fit distance. The likelihood function gives the probability of

139   observing the data given the parameters. Maximizing the likelihood implies selecting the

140   parameters for which the probability of observing the data is highest. Maximum likelihood is by

141   far the most standard approach to distribution fitting and more generally to model fitting. It is

142   backed with a consequent body of theoretical work ensuring many interesting asymptotic

143   properties[18]: the maximum likelihood estimate converges to the true value of the parameters

144   (consistency), it is the fastest estimate to converge (efficiency), the difference with the true value

145   is normally distributed (normality), which provides confidence intervals on the parameters. This

146   approach is used by the Australian software Burrlioz[19], for instance, which fits a Burr III

147 distribution to the toxicity data. Moreover, a natural extension of the likelihood function makes it

148 possible to take into account censored data[20].

149      Another popular method to determine the distribution parameters that best describe the

150 data is the least-square regression on the empirical Cumulative Distribution Function (CDF),

151 which minimizes the sum of the squared vertical distances between the CDF and the data. It is

152 also known as the Cramer-Von Mises distance[21]. This approach is adopted by the software

153 CADDIS_SSD[22], from US EPA, which performs a least-square regression on the CDF with a

154 log-probit function. The software SSD MASTER[23] uses the same method, but tries several

155 other distributions: normal, logistic, Gompertz and Fisher-Tippett. However, regression on the

156 CDF is not an easy method to include censored data, because constructing the CDF implies

157 sorting the data, which is not trivial with censored data. In addition, there is no unique way to

158 build a CDF. Several possible plotting positions[1,24] all have desirable properties, but none of

159 them represent the data more faithfully than any other. Therefore, the resulting SSD and its

160 predictions depend on purely arbitrary decisions regarding the plotting positions, a fact that

161 undermines its scientific credibility.

162      A third common approach to determine the distribution parameters is to match the

163 moments of the empirical distribution with those of the model. This is numerically easy when

164 there is an analytical formula for the determination of the parameters. Moment matching is

165 equivalent to maximum likelihood for the distributions of the exponential family, such as the

166 normal, exponential, gamma distributions, but not for the logistic distribution. ETX[25] is a free

167 software which uses the moment matching method[14,15]. It fits a log-normal and a log-logistic

168 distribution by moment matching and computes confidence intervals. Moment matching is

169  sensitive to outliers and can give unrealistic results[26]. However, it can be useful when the

170  maximum likelihood computation is intractable[26]. However, moment determination with

171  censored data is not trivial. Therefore, this method cannot be straightforwardly extended to

172  include censored data.

173  This brief review of the classical SSD approaches shows that several methodological

174  choices must be made in order to fit an SSD: whether or not to use a non-parametric method,

175  selection of the distribution and of the parameter estimation method. Apart from maximum

176  likelihood, there is no straightforward approach for a non expert in statistics to make use of all

177  types of censored data. Indeed, all of the available turn-key software for SSD fitting require the

178  use of non censored data. Yet, there is a possibility to use the R-package *fitdistrplus[10]* to fit

179  censored data using maximum likelihood, with the following scheme.

180

## Maximum likelihood for censored data

182  Maximum likelihood provides a single framework to cope with both censored and non-

183  censored data. Censored data is a general name given to data which are not in the form of fixed

184  values but belong to an interval, bounded or not. Censored sensitivity data occur when it is not

185  possible to determine a CEC for a given species. Possible reasons are 1) that the highest

186  concentration tested does not have any noticeable effect, 2) that only a tiny amount of

187  contaminant already stamps out all the individuals, 3) that the measurement is simply too

188  imprecise to be reasonably described by a single value instead of an interval. In such cases, it is

189  only possible to give a *lower* bound, a *higher* bound or an *interval* to the CEC. Such data are

190  called *right-censored*, *left-censored* or *interval-censored*, respectively. Censored data can also occur when there are multiple values for the sensitivity of one species to a given toxicant. When the quality of the data seem equivalent, ECHA's advice[5] are to use the geometric mean as a replacement for the different values. It might be more cautious to use these multiple values to define an interval containing the sensitivity of that species. Censored data are very different from doubtful or questionable data, obtained from failed experiments. They are produced using a valid experimental procedure and they contain information as valid as non-censored data. Censorship is very common, especially for rare species where there are scant data available and for which no standard test procedure exists. There is a downside in discarding censored data, as they could represent the better part of an extended dataset. For instance, in the work by Dowse et al.[8], discarding censored data entails a division of the number of tested species by a factor 8.

In spite of their ubiquity, censored data appear to be very much ignored in ecotoxicology. To our knowledge, there is no example of SSD including all types of censored data in a frequentist framework. It is possible in a Bayesian framework[24,27,28], but fitting a Bayesian model requires a certain statistical expertise. Censored data are typically either discarded or substituted with arbitrary values, which is a bias-prone approach in general[9]. However, there is a simple method to include censored data in a frequentist framework. Parameter estimation of a distribution on any type of censored data can be performed using a natural extension of the maximum likelihood method[29]. Let $x_i$ be $N$ sensitivity data following distribution $f$ of parameter $\theta$. The likelihood function for non-censored data writes as follows:

$$L(\theta) = \prod_{i=1}^{N} f(x_i|\theta)$$

(1)

This likelihood function can be extended to censored data. Let $x_i$ be the $N_{nc}$ non-censored data,

212 $x_j^{up}$ the $N_{lc}$ upper bounds for left-censored data, $x_k^{low}$ the $N_{rc}$ lower bounds for right-censored

213 data and $(x_l^{low}, x_l^{up})$ the $N_{ic}$ pairs of bounds for interval-censored data. Then, the previous

214 likelihood function is now extended to:

$$L(\theta) = \prod_{i=1}^{N_{nc}} f(x_i|\theta) \times \prod_{j=1}^{N_{lc}} \left(F(x_j^{up}|\theta)\right) \times \prod_{k=1}^{N_{rc}} \left(1 - F(x_k^{low}|\theta)\right) \times \prod_{l=1}^{N_{ic}} \left(F(x_l^{up}|\theta) - F(x_l^{low}|\theta)\right)$$

215 (2)

216 where $F$ is the cumulative distribution function of distribution $f$.

217 We see that the likelihood function for censored data (eq. 2) writes as a product of four terms, the

218 first being the likelihood for non-censored data (eq. 1) and the next three corresponding to the

219 left-censored data, right-censored data and interval-censored data respectively.

## MOSAIC_SSD

221 It is possible to use the method described in the previous section using the R-package

222 *fitdistrplus[10]*. R-packages *survival[30]* and *NADA[31]* offer the same possibility. However,

223 they require a certain fluency in the R programming language, preventing the widespread use of

224 censored data in ecotoxicology. Minitab[32] is a commercial software with a graphical user

225 interface which fits multiple distributions to censored data rather easily, but there does not seem

226 to be any open-source alternative.

227 Moreover, *fitdistrplus* and these other packages and software are not specifically designed for

228 SSD and their versatility in the choice of distributions and fitting methods might discourage

229 inexperienced users. Thus, we developed a web-interface, MOSAIC_SSD (http://pbil.univ-

230 lyon1.fr/software/mosaic/ssd/), which is a wrap up of *fitdistrplus* into a SSD-dedicated online

231 tool. MOSAIC_SSD enables anyone to perform a simple, yet statistically sound SSD analysis

232 including censored data without worrying about the conceptually difficult underlying statistical

233 questions. The web interface is easily accessible via any browser and simple to use: given an

234 input dataset, it sends the calculation to a server then hands in the result. The input dataset is a

235 text file uploaded by the user with a straightforward encoding. A non censored dataset is given in

236 one column. A censored dataset is given as two columns: a "NA" in the right – resp. left –

237 column and a number on the left – resp. right – denotes a right – resp. left – censored data. Two

238 differing numbers denote an interval-censored data and two identical numbers a non-censored

239 data.

240     Few options are offered to keep the tool more user-friendly. The user can choose one or

241 two among the log-normal and log-logistic distributions. These two distributions are the most

242 widely used[16], and parameter estimation appears robust enough to accommodate for most

243 datasets, as they contain only two parameters. In order to select which distribution describes the

244 data best, the first step is to perform a qualitative assessment by looking at the representative

245 curves. The value of the likelihood function for each model can then be used as a further decision

246 criterion. The log-logistic distribution has heavier tails than the log-normal and is therefore more

247 conservative in the determination of the $\mathrm{HC}_5$[15].

248     The second choice left to the user is to decide whether to compute the bootstrap 95%

249 confidence intervals. The calculation runs slightly longer with the bootstrap but it yields

250 confidence intervals on the parameters of the distribution and on several computed

251 $\mathrm{HC}_p$ $(p = 5, 10, 20, 50)$. The bootstrap procedure is not guaranteed to converge, the number of

252 iterations required being strongly dependent on the dataset. Therefore, an automatic check of

253 bootstrap convergence is implemented. The procedure is run several times in parallel, comparing

254    the magnitude of the results fluctuations to the span of the confidence interval. This comparison

255    determines whether the bootstrap has converged. In the case were the bootstrap procedure fails to

256    converge, additional computations are launched. If the bootstrap finally converges, or if the

257    process has reached the time limit, the user is advised whether the confidence intervals are

258    reliable. Calculating the confidence intervals using a bootstrap method has the advantage of

259    using a unified framework for every distribution. Figure 1 shows a screenshot of the result page

260    of the analysis with an example dataset (provided in MOSAIC_SSD) containing censored data

261    and documenting the salinity tolerance of riverine macro-invertebrates[33] (hereinafter referred

262    to as the *censored salinity* dataset). The dataset contains 72-hrs $LC_{50}$ values (lethal concentration

263    for 50% of the organisms) for 110 macro-invertebrate species from Australia. Data were

264    collected using rapid toxicity testing[34] and contain non-censored, right-censored and interval-

265    censored data. The result page shows a graphical representation of the censored data, the

266    distribution parameters, $HC_p$ computed for various interesting values of $p$ and the bootstrap

267    confidence intervals within brackets. Figure 1 also shows the output of an SSD analysis with a

268    non censored dataset. It actually is a non-censored version of the salinity dataset described

269    earlier. The transformation from censored to non-censored dataset follows the customary

270    approach to censored data, which consists in discarding some type of data and transforming

271    others (more details in the next section). An analysis with non censored data follows identical

272    steps and yields results with the same outline, except that a traditional CDF is used to visualise

273    the data. The obvious difference between the outputs of the censored dataset and the non-

274    censored dataset is the representation of the CDF. For non-censored data, the CDF is represented

275    using the traditional Hazen plotting positions[1]. The choice of plotting positions remains

276     arbitrary and there is no perfect solution[1,24], so preference was given to the most standard

277     approach. Representation of censored data CDF is far from evident. Building a CDF implies

278     defining an ordering for the data. If obvious for non-censored data, such an ordering makes little

279     sense for interval-censored data. They might be ranked according to the median of the interval, to

280     the higher bound or the lower. Adding left or right-censored data complicates matter even more.

281     Within *fitdistrplus*, the answer to this problem is to use the Turnbull estimate of the CDF, which

282     is a non-parametric maximum likelihood estimator of the CDF[35]. This estimate can be

283     obtained through an expectation-maximisation algorithm and yields the CDF which predicts the

284     data with the highest probability. The Turnbull estimate is represented as a stepwise curve as on

285     Figure 1 (top panel).

286          Finally, MOSAIC_SSD can be used as a stepping stone to perform further analysis with

287     *fitdistrplus*. The last item on the MOSAIC_SSD result page is not shown on the screenshots. It is

288     an R script offering the possibility to replicate the analysis using *fitdistrplus,* through a copy and

289     paste operation in R. This script is intended as a stepping stone to using the complete *fitdistrplus*

290     R-package. It can be adjusted by slightly changing some of the options. For instance, $HC_p$ for

291     different values of $p$ can be computed, with an alternative distribution or a different fitting

292     method. Moreover, this script ensures transparency and traceability of the results obtained

293     through MOSAIC_SSD.

294

295

296

297

**Added value of including censored data**

298

299    Changing a few parameters in the R script provided within MOSAIC_SSD, it is possible

300    to use *fitdistrplus* to investigate on several fundamental aspects of SSD, such as the influence of

301    including censored data on the prediction. A customary approach when dealing with censored

302    data is to discard or to transform it. More precisely, it is frequent to discard left or right-censored

303    data and to take the middle of the interval-censored data as a single value. Two datasets were

304    analysed to assess the effect of such data transformation on the predicted hazardous

305    concentrations. In the censored salinity dataset mentioned earlier, out of 108 $LC_{50}$, 89 (82.4%)

306    are censored, among which 60 (55.6%) are right-censored and 29 (26.8%) interval-censored.

307    Most of the censored data resulted from the testing of rare species, for which the small number of

308    individuals captured prevented the calculation of an $LC_{50}$ by fitting a concentration-effect

309    model[33]. This extensive dataset was collected to be as representative as possible of the species

310    found in nature[33]. Therefore, a first asset of taking censored data into account is to abstain

311    from discarding or altering the vast majority of the data. The resulting SSD is therefore more

312    representative of the community it aims to describe. Moreover, using only non-censored data in

313    the analysis introduces a strong selection bias towards abundant species. This is particularly

314    problematic, when some rare species are likely to be among those that the environmental

315    manager wishes to protect by carrying out an SSD analysis. The second dataset was published by

316    Koyama et al.[36], and contains vertebral deformity susceptibilities of marine fishes exposed to

317    trifluralin (hereinafter referred to as the *censored trifluralin* dataset). The measured endpoint are

318    96-hrs $LC_{50}$ on 10 species. Four of the $LC_{50}$ are censored, among which two are right-censored

319    and two are left-censored. On this dataset, the obvious advantage of taking censored data into

320 account is that the SSD can be fitted on 10 species, whereas discarding the censored data reduces

321 the size of the dataset to six species only. This is below the minimum recommendation of ECHA

322 (of 10, preferably 15[37]). A non-censored version of the two datasets (hereinafter referred to as

323 the *transformed salinity* and *transformed trifluralin* datasets) was obtained following the habitual

324 procedure of discarding the right or left-censored data and taking the middle of the interval-

325 censored data. Fitting the lognormal distribution on the censored and transformed versions of the

326 datasets showed that discarding censored data had an adverse effect on the predicted $HC_5$

327 (Figure 2). For the salinity dataset, discarding left-censored data induced a clear upward bias for

328 the cumulative curve and a therefore greater $HC_5$ (Figure 2 left). The estimates for the $HC_5$

329 were: $9.85 \ \text{g.L}^{-1}[8.38; 11.80]$ for the censored dataset and $7.98 \ \text{g.L}^{-1}[6.63; 9.93]$ for the

330 transformed dataset, respectively. An unnecessary high hazardous concentration might seem a

331 harmless error, since it is more protective to use the transformed salinity dataset. However, that

332 incorrectly low value might motivate the use of costly decontamination measures at a specific

333 location, when efforts could be spared and distributed elsewhere.

334 The influence of censored data is dataset-dependent and the bias could be in the opposite

335 direction. This is illustrated on the trifluralin dataset (Figure 2 right). Fitting the log-normal

336 distribution yielded the following estimates for the $HC_5$:

337 $2.4 \times 10^{-3} \text{mg.L}^{-1}[4.7 \times 10^{-5}; 2.6 \times 10^{-2}]$ for the censored dataset and

338 $1.7 \times 10^{-2} \text{mg.L}^{-1}[8.9 \times 10^{-3}; 4.3 \times 10^{-2}]$ for the transformed dataset, respectively.

339 Discarding the censored data led to underestimate the variability in the community sampled by

340 the tested species, resulting in a smaller $HC_5$. Therefore, the width of the distribution was

341  underestimated and the fifth percentile had a larger value. On the *trifluralin* dataset, discarding

342  the censored data led to an underestimation of the trifluralin real toxicity and its potential hazard

343  to the environment. Another striking differentiation was that the span of the confidence interval

344  was much larger when censored data were included in the SSD. It reveals that a possible effect of

345  transforming censored datasets is to severely underestimate the width of the confidence interval

346  and to give overconfident predictions on the hazardous concentrations.

347  **Discussion**

348      In this paper, we reviewed the general approach to fit an SSD to sensitivity data and

349  explained how it was possible to use maximum likelihood to include censored data in SSD. We

350  presented MOSAIC_SSD, a web-tool which enables any user to perform an SSD analysis

351  including censored data with few very simple steps. MOSAIC_SSD is an interface to a more

352  versatile tool, the R package *fitdistrplus[10]* and presents only few options to simplify the use.

353  We supported the methodological approach behind MOSAIC_SSD with several arguments and

354  showed the added value of including censored data into the SSD. Discarding or transforming

355  censored data has been shown to alter the results of the SSD analysis. Using MOSAIC_SSD is a

356  convenient way to take censored data into account in the fitting of an SSD. Moreover, the sound

357  general statistical approach is also an asset to perform any sort of SSD. Considering the choice of

358  a distribution, MOSAIC_SSD provides by default two standard distributions, the log-normal and

359  log-logistic, but it encourages the use of alternative distributions by providing a stepping stone to

360  using the R package *fitdistrplus*. The question "which distribution best fits a dataset?" cannot

361  have a general answer and must be addressed by testing several options. Therefore, the

362  possibility to try multiple distributions is a valuable asset. For instance, a user might wish to fit a

363 distribution that best describes the *tails* of the dataset, because determining a $HC_5$ is an extreme

364 quantile estimation problem. In that case, a heavy tailed distribution such as Weibull or

365 exponential is appropriate.

366     In selecting a distribution, it is important not to pick a distribution with too many

367 parameters. One of the easily accessible software for SSD is BurrliOZ[19], which fits the Burr

368 III distribution using maximum likelihood and computes confidence intervals using bootstrap.

369 The Burr III distribution is very flexible[13], but it contains one parameter more than the log-

370 normal or log-logistic distributions. Fitting of a distribution with many parameters requires a lot

371 of data and the Burr III distribution is likely to suffer from strong structural correlation among

372 the parameters[13]. Therefore, convergence can be difficult and the estimates produced are not

373 very reliable. However, BurrliOZ is currently being developed to fit the log-logistic distribution

374 on small datasets and to provide a comparison between at least the log-logistic and the Burr III

375 distribution for larger datasets[3].

376     MOSAIC_SSD, easily accessible and user-friendly, can encourage the inclusion of

377 censored data in SSD analysis in order to better use all the data at hand.

378 We did not address all the methodological issues relating to the SSD approach but tried to

379 improve the existing methods, with the aim to make the most of the available data given the cost

380 of collecting them. There remain interrogations as to what might happen if the proportion of

381 censored data is too great and the dataset is small. It is not possible to test this situation

382 thoroughly, for there are many ways to censor data and no trivial way to choose between them. A

383 good practice would be to consider the span of the confidence interval around the hazardous

384 concentration of interest and decide if the dataset is adequate for predicting such concentrations

385 or if more data need to be collected. Taking censored data into account would therefore be

386 crucial to have a precise assessment of the confidence interval, and not an artificially reduced

387 estimation as in the trifluralin dataset.

388       We mentioned that censored data might represent an important part of any dataset and

389 that MOSAIC_SSD could be profitably used on many occasions. However, this work could have

390 a more general scope, since fundamentally all data with a confidence interval could be

391 considered as interval-censored data. Indeed, the confidence interval around an $EC_{50}$ or any

392 CEC estimate can be considered as the range which has a 95% probability of containing the real

393 value and be reported as an interval-censored data. Using the confidence intervals on the CECs

394 as censored data provides a basic way to propagate the uncertainty on the CEC into the SSD, a

395 fundamental problem of SSD[3] which is seldom addressed[38].

396 Moreover, LOEC (lowest observed effect concentration) data, which are often reported, are

397 indeed left-censored data. The only information LOEC carries is that the NOEC lies below this

398 concentration[39]. Therefore, the SSD approach we propose, which includes censored data,

399 would allow ecotoxicologists to better use the available experimental data used to calculate the

400 NOEC.

401       However, we reached the limits of a traditional SSD based on CEC's and still discarded a

402 lot of information. Indeed, a CEC is only a summary of a full concentration-effect curve. This

403 summary sets aside several aspects of the response of a species to a pollutant, such as the slope

404 of the curve. This slopes describes whether the species is gradually affected or there is a sudden

405 drastic effect. It is possible to include all the information present in the experimental

406 concentration-effect curve in the SSD by building a hierarchical model of SSD. This hierarchy would model the joint probability of all the parameters describing a concentration-effect curve, not only the CEC in the classical SSD[40]. Moreover, this would also take proper account of the uncertainty on the species response modelling and to propagate uncertainty into the SSD.

## Acknowledgements

Financial support for the PhD of Guillaume Kon Kam King is provided by the Région Rhône-Alpes. The authors thank Cécile Poix for English proofreading.

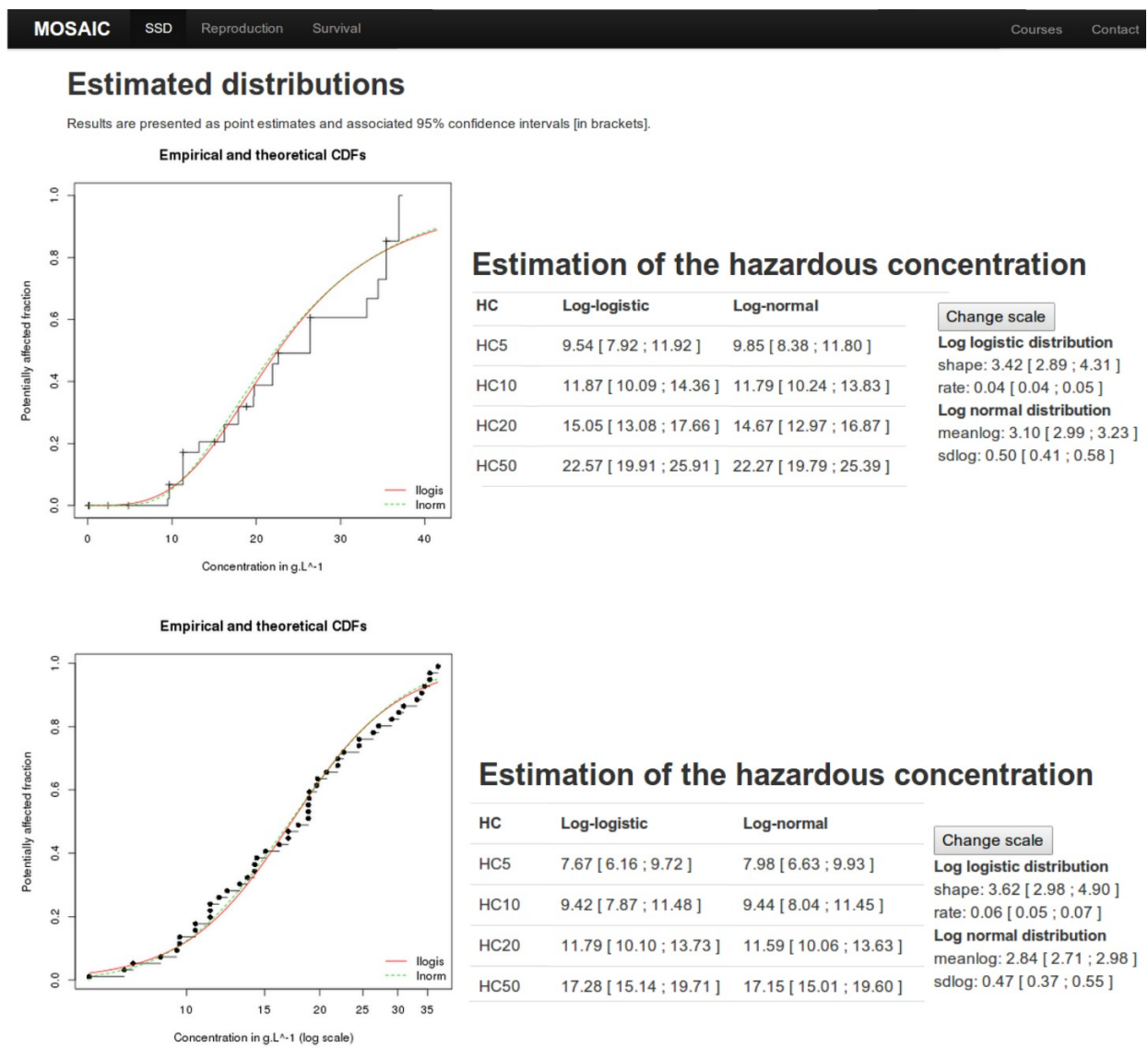

Figure 1: Screenshot of the result page of MOSAIC_SSD on the salinity censored dataset. Top shows the output of MOSAIC_SSD on the salinity dataset, bottom shows the output on a non-censored dataset obtained from a transformation of the salinity dataset.

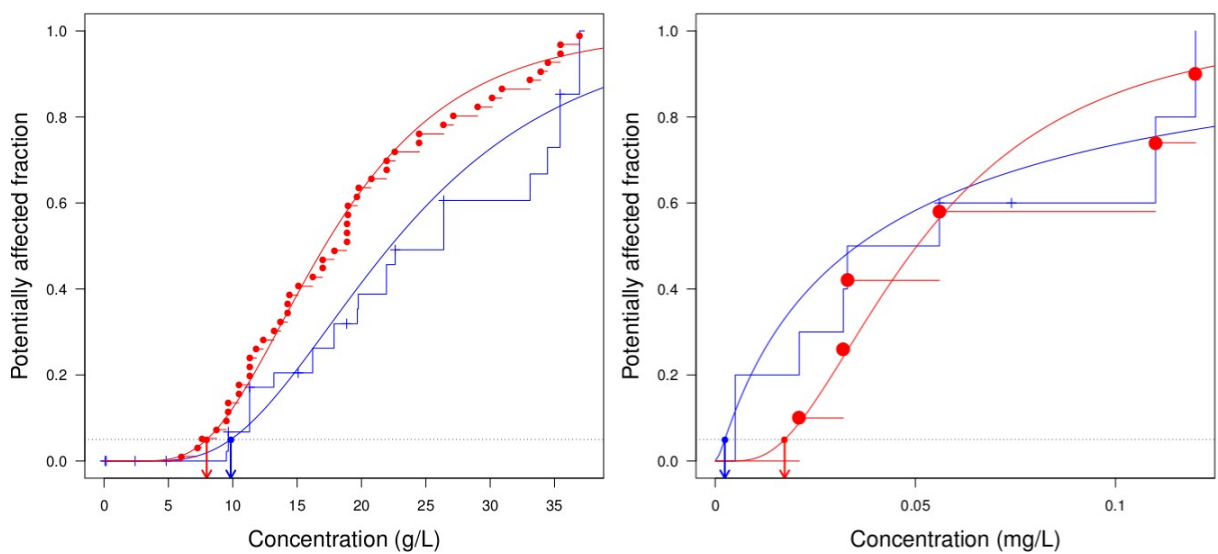

Figure 2: Fitted and empirical cumulative distribution and the $HC_5$ for the salinity dataset (left) and the trifluralin dataset (right). The dotted line corresponds to a potentially affected fraction of 5%. Vertical arrows indicate the $HC_5$. Only the region around the $HC_5$ is represented. The blue line is for the censored dataset, the red for the transformed dataset. A log-normal distribution was fitted on both datasets.